\documentclass[aps,amsmath,amssymb,12pt]{revtex4}
\usepackage{graphicx}
\usepackage{bm}

\def\journal#1#2#3#4{{#1} {\bf #2}, #3 (#4)}
\newcommand{\be}{\begin{equation}}
\newcommand{\ee}{\end{equation}}
\newcommand{\bea}{\begin{eqnarray}}
\newcommand{\eea}{\end{eqnarray}}
\newcommand{\hf}{\frac12}
\newcommand{\nn}{\nonumber\\}
\def\eq#1{(\ref{#1})}
\def\la{\langle}
\def\ra{\rangle}

\def\Tr{{\mathrm{Tr}}}
\def\ord#1{{\cal O}\left(#1\right)}

\def\v#1{{\bm{#1}}}
\def\fd#1#2{\frac{\delta#1}{\delta#2}}
\def\fdd#1#2#3{\frac{\delta^2#1}{\delta#2\delta#3}}
\def\dk{\Delta k}

\def\hD{{\hat D}}

\def\t{\tilde}

\begin{document}
\title{Euclidean scalar field theory in the bi-local approximation}

\author{S. Nagy$^1$, J. Polonyi$^2$, I. Steib$^1$}
\affiliation{$^1$Department of Theoretical Physics, University of Debrecen,
P.O. Box 5, H-4010 Debrecen, Hungary}
\affiliation{$^2$Strasbourg University, CNRS-IPHC, \\23 rue du Loess, BP28 67037 Strasbourg Cedex 2, France}
\date{\today}

\begin{abstract}
The blocking step of the renormalization group method is usually carried out by restricting it to fluctuations and to local blocked action. The tree-level, bi-local saddle point contribution to the blocking, defined by the infinitesimal decrease of the sharp cutoff in momentum space, is followed within the three dimensional Euclidean $\phi^6$ model in this work. The phase structure is changed, new phases and relevant operators are found and certain universality classes are restricted by the bi-local saddle point.
\end{abstract}

\maketitle

\section{Introduction}
The realistic physical models involve a large number of particles. The best method to simplify the models is the successive elimination of the less important degrees of freedom, the renormalization group (RG) method. One introduces here a gliding cutoff and eliminates the particle modes beyond the cutoff. This step, called blocking, generates a new, rather complicated blocked action which is usually simplified by employing the loop-expansion in the elimination and by ignoring the resulting non-local terms. The dependence of the coupling constants on the cutoff is defined by the evolution equations, usually restricted to the fluctuations, assuming tacitly the absence of tree-level contributions with the remarkable exception of vacuum with spontaneously broken symmetry \cite{tree}. It has been noted recently that independently of the symmetry breaking the blocking always introduces another tree-level contribution which is bi-local \cite{ctprg}. In the present work a more detailed account is given about the impact of such saddle point on the phase structure and the RG flow of a three-dimensional Euclidean scalar model. The fact that these terms have been overlooked for a long time might be the result of their bi-local and tree-level nature. 

The physical reason to expect such contributions to the blocking is rather obvious: an important tree-level phenomenon of the eliminated modes is the radiation and the radiation field, being a mass-shell effect, is not accessible within the framework of the gradient expansion, the usual approximation method to treat the non-local effective dynamics. In fact, while the latter is based on the expansion in the momentum space around the origin, the former displays a singular dependence on the momentum. Hence some conclusions of the RG studies, traditionally including the local loop contributions only, should be reconsidered by allowing tree-level, non-local blocked action. The present work, treating a simple, Euclidean scalar field theory, is a modest step in that direction.

The tree-level contributions to the RG evolution equation has already been considered in the broken symmetric phase of the scalar model and it was found to be the origin a generalized Maxwell-construction, the degeneracy of the action within the mixed phase \cite{tree}. The corresponding saddle point is a plane wave and has zero modes which ares supposed to restore the space-time symmetry of the vacuum. The goal of the present work is to explore a different, simpler saddle points of the blocking, without zero modes. Another novelty, compared to ref. \cite{ctprg}, is the taking into account the loop corrections to the RG flow.

Non-local theories have been present on the landscape of field theory for a long time. First it has been hoped that theories with non-local action have UV finite perturbation expansion \cite{uvfinite,krizhnits}. However such theories are plagued by serious problems, the violation of the causality \cite{causality}, the spoiling of gauge invariance \cite{gauge}, and the emergence of instability \cite{instab,eliezer}. Finally they gained a general acceptance by realizing that a regulated theory is always non-local at the scale of the cutoff. Rather than continuing the exploration of the problems of realistic, real time theories we restrict our attention to the modification of the RG flow, due to some simple bi-local terms in the action, not being related to regulators.

The RG evolution equation is a functional differential equation whose solution is beyond our analytical capabilities. To obtain a soluble scheme one usually projects the evolution equation into a functional subspace of the effective action, defined by an ansatz. The perturbative elimination of the modes generates multi-local effective action, the $n$-cluster contribution being an $n$-fold space(-time) integral. Our ansatz restricts this structure to $n\le2$, i.e. to local and bi-local contributions and simplifies the field dependence to $\ord{\phi^6}$. We explore the properties of the bi-local potential in the framework of the Wegner-Houghton equation \cite{wh}. Our choice of the RG scheme is motivated by the need of an analytical expression of the saddle point, available for sharp cut-off only. The effective action based renormalization with smooth cut-offs \cite{wetterich,morris} does not limit the volume of the eliminated modes in the momentum space, needed to simplify the saddle point. 

An extension of the ansatz of the effective action leads several differences in comparing the local and a bi-local theories, namely new relevant operators are generated at the Gaussian fixed point, a partial non-Gaussian fixed point appears where only the local coupling strengths remain constant, new, modulated phase is found and an instability restricts the available parameter space of the theory. Furthermore a change in the universality class structure of the $\phi^4$ model, defined before or after the thermodynamic limit, indicates that the thermodynamic limit of the momentum space regulated theory does not always converge to the theory, defined in the strictly infinite volume. While the bi-local terms may lead to a rich phase diagram compared to the local theory this direction is not explored below where the initial conditions for the RG equation are restricted to a local action. 

The presentation starts with the introduction of the ansatz for the blocked action in Sec. \ref{ansatzs}, followed by Sec. \ref{sec:evol} with the derivation of the evolution equation. The phase diagram and the RG trajectory flow are presented in Sec. \ref{sec:dis}. Sec. \ref{stabs} is devoted to the discussion of the stability and the universality. Finally, the discussion of our results is presented in Sec. \ref{sums}.

\section{Bi-local action}\label{ansatzs}
The effective, blocked action is assumed to be the sum, $S=S_0+S_1+S_2$, of a free action, 
\be
S_0=\hf\int_x(\partial\phi_x)^2,
\ee
$\int_x=\int d^dx$, a local
\be
S_1=\int_x\tilde U(\phi_x),
\ee
and a bi-local term,
\be
S_2=\int_{xy}\tilde V_{x-y}(\phi_x,\phi_y).
\ee
The symmetry under the discrete internal space transformation, $\phi\to-\phi$, is assumed, $\tilde U(\phi)=\tilde U(-\phi)$ and $\tilde V_{x-y}(\phi_1,\phi_2)=\tilde V_{x-y}(-\phi_1,-\phi_2)$. The bi-local potential is supposed to be symmetric with respect to $O(d)$ rotations, $V_{x-y}=V_{|x-y|}$, and to the exchange of the cluster locations, $\tilde V_{x-y}(\phi_1,\phi_2)=\tilde V_{y-x}(\phi_2,\phi_1)$ and its Fourier expanded form,
\be
\tilde V_q(\phi_1,\phi_2)=\int_xe^{iq(x-y)}\tilde V_{x-y}(\phi_1,\phi_2),
\ee
will frequently be used with the convention that $x-y$ and $q$ in its index denote the variable of the translation invariant function in the coordinate and in the Fourier space, respectively.

It is more illuminating to display the RG equations by allowing unrestricted field dependence in the potentials hence a more compact form of the evolution equation is derived in that form, to be truncated in the numerical work to a polynomial of finite order. By anticipating a homogeneous condensate in the first stage of the calculation we split the field, $\phi_x=\Phi+\chi_x$ where $\int_x\chi_x=0$, and remake the separation of the local and the bi-local terms in the action, 
\be
S=L^dW+S_0[\chi]+\int_xU(\chi_x)+\int_{xy}V_{x-y}(\chi_x,\chi_y),
\ee
where the $\Phi$-dependence is suppressed on the right hand side and
\bea
U(\chi)&=&\tilde U(\Phi+\chi)-\tilde U(\Phi)+\tilde V_0(\Phi+\chi,\Phi)+\tilde V_0(\Phi,\Phi+\chi)-2\tilde V_0(\Phi,\Phi),\nn
V_{x-y}(\chi_1,\chi_2)&=&\tilde V_{x-y}(\Phi+\chi_1,\Phi+\chi_2)-\tilde V_{x-y}(\Phi,\Phi+\chi_2)-\tilde V_{x-y}(\Phi+\chi_1,\Phi)+\tilde V_{x-y}(\Phi,\Phi),\nn
W&=&\tilde U(\Phi)+\tilde V_0(\Phi,\Phi).
\eea
The functions, appearing in our ansatz, are assumed to be analytic in the field,
\bea\label{polynans}
U(\chi)&=&\sum_{n=2}^N\frac{g_n}{n!}\chi^n,\nn
V_{x-y}(\chi_1,\chi_2)&=&\sum_{m+n=2}^N\frac{V_{x-y,m,n}}{m!n!}\chi_1^m\chi_2^n,
\eea
the value $N=6$ being used in the numerical work in dimension $d=3$.

We shall need later the first two functional derivatives,
\bea
\fd{S}{\chi_x}&=&-\Box\chi_x+U'(\chi_x)+\int_y\partial_1V_{x-y}(\chi_x,\chi_y)+\int_y\partial_2V_{y-x}(\chi_y,\chi_x),\nn
\fdd{S}{\chi_x}{\chi_y}&=&-D^{-1}_{xy}[\chi]=D^{-1}_{xy}-\Sigma_{xy}[\chi].
\eea
The inverse propagator on the homogeneous field, $D^{-1}_q=q^2+K_q+m^2$, is given by the help of the mass square, $m^2=U''(0)+(\partial_1^2+\partial_2^2)V_0(0,0)$, and the correction $K_q=2\partial_1\partial_2V_q(0,0)$ to the kinetic energy where the notation $\partial_jf(\chi_1,\chi_2)=\partial_{\phi_j}f(\chi_1,\chi_2)$ is used. The self energy assumes the form,
\bea
\Sigma_{xy}[\chi]&=&-\delta_{xy}\Delta_\Phi U''(\chi_x)-2\Delta_\Phi\partial_1\partial_2V_{x-y}(\chi_x,\chi_y)\nn
&&-\delta_{xy}\int_z[\Delta_\Phi\partial^2_1V_{x-z}(\chi_x,\chi_z)+\Delta_\Phi\partial^2_2V_{z-x}(\chi_z,\chi_x)],
\eea
by using the notation $\Delta_\Phi f(\chi)=f(\Phi+\chi)-f(\Phi)$.

\section{The evolution equation}\label{sec:evol}
The functional integral over the modes to be eliminated is usually evaluated by the help of the saddle point expansion. The saddle point is rather involved when a smooth cutoff is used and to simplify the matters we use a sharp cutoff, $|\v{p}|<k$, for the blocking, consisting of the decrease, $k\to k-\dk$, of the gliding cutoff. The saddle point simplifies and the higher loop contributions become suppressed for small $\dk/k$ hence the decrease of the cutoff is chosen to be infinitesimal. This is possibly in the thermodynamic limit, assumed to be performed before starting the calculation. The field variable is splitted into the sum, $\phi\to\phi+\varphi$, where $\phi$ ($\varphi$) denotes the IR (UV) component, being non-vanishing for $0<|p|<k-\dk$ ($k-\dk<|p|<k$), and the blocked action is given by
\be\label{blocking}
e^{-\frac1\hbar S_{k-\dk}(\phi)}=\int D[\varphi]e^{-\frac1\hbar S_k[\phi+\varphi]}.
\ee

It is instructive to write the local part of the action on the right hand side in the form
\be
S^{loc}_k[\phi+\varphi]=S_k[\phi]+S_0[\varphi]+\sum_n\frac1{n!}\prod_{j=1}^n\int_{p_j}\varphi_{p_j}{\cal U}_{p_1,\ldots,p_n}
\ee
where $\int_p=\int\frac{d^dp}{(2\pi)^d}$ and
\be
{\cal U}_{p_1,\ldots,p_n}=\sum^{N_{loc}}_{m=n}\frac{g_m}{(m-n)!}\prod_{j=n+1}^m\int_{q_j}\phi_{q_j}\delta_{0,\sum_jq_j+\sum_\ell p_\ell}.
\ee
The perturbative evaluation of the functional integral \eq{blocking} generates a change $\Delta_kS_k[\phi]$, expressed by help of the notation $\Delta_k f_k=f_k-f_{k-\dk}$, as the sum of connected graphs where the internal lines represent the free UV propagator, $\Theta(|p|-k-\dk)\Theta(k-|p|)D_p$, and the external legs belong to the IR field attached to ${\cal U}$. To simplify the momentum integrals it is assumed that the IR field has a continuous Fourier transform. The $\ord{\hbar^0}$ part of the evolution, generated by the local part of the action, is given by tree graphs and each internal line comes with a factor $\dk$ due to the continuity of $\phi_p$. The real-space form,
\be
S^{loc}[\phi+\varphi]=S_0[\phi]+S_1[\phi]+\sum_n\frac1{n!}\int_x\varphi^n_x\tilde U^{(n)}(\phi_x),
\ee
reveals that each factor ${\cal U}$ generates a local $\phi$-cluster and the cluster contributions beyond the bi-local level are suppressed in the limit $\dk\to0$.

The small parameter, organizing the loop expansion of the functional integral \eq{blocking} is $\hbar\dk/k$ hence it is sufficient to consider the tree and the one-loop contributions,
\be\label{discrev}
e^{-\frac1\hbar S_{k-\dk}[\phi]}=\int d\alpha e^{-\frac1\hbar S_k[\phi+\varphi^s_\alpha(\phi)]-\hf\Tr\ln\fdd{S_k[\phi+\varphi^s_\alpha(\phi)+\varphi]}{\varphi}{\varphi}_{|\varphi=0}},
\ee
where $\varphi^s_\alpha$ denotes the saddle point, corresponding to the zero mode parameter, $\alpha$. A word of caution is needed here about the zero mode integration, it is not carried out if $\alpha$ influences the saddle point in an infinitely large volume. In fact, local excitations over such a vacuum preserve $\alpha$, characterizing equivalent, degenerate vacuums, an example being the modulated vacuum of the density wave phase of solids \cite{gruner}. The $k$-dependence, shown explicitly in these equations, will be suppressed below if no confusion can arise. The action at the initial cutoff, $k=\Lambda$, is assumed to be local, $V=0$, for the sake of simplicity.

\subsection{Tree-level evolution}\label{treeevs}
One has to treat saddle points with and without zero modes separately since the zero modes display large fluctuations in the absence of the restoring force to an equilibrium position, generating important changes in the dynamics. We work with a small but finite $\dk$ and the saddle point is spread over the distances $1/\dk$, letting the zero modes of the saddle point of the blocking integrated over. The UV modes are inhomogeneous by construction, the saddle point breaks the space-time symmetries and some of the possible zero modes are the corresponding Goldstone modes. 

The saddle point minimizes the action, $S_k[\phi+\varphi^s]$, a problem reminiscent of the variational method to obtain an approximate ground state wave function in quantum mechanics. The restriction to the thin shell of the UV modes in the momentum space renders the problem similar to the calculation of the low lying collective modes in a fermi liquid. In fact, the periodic potential takes the role of the ionic crystal and the variational ansatz,
\be\label{singlsp}
\varphi^s_{n,\theta,q}=\frac\rho2(2\pi)^d\left[e^{i\theta}\delta\left(q-n\left(k-\frac\dk2\right)\right)+e^{-i\theta}\delta\left(q+n\left(k-\frac\dk2\right)\right)\right],
\ee
where phase, $\theta$, and the direction of the wave vector, $n$, $n^2=1$, are zero mode parameters for $\chi=0$ \cite{tree} is an analogy of the density waves due to a nested Fermi-surface \cite{gruner}. The resulting renormalized trajectory converges in the limit $\dk\to0$ and realizes a generalized Maxwell-cut, rendering $S_{k-\dk}$ degenerate up to a variation $\ord\dk$ within the domain where $S_k$ is concave \cite{vincent}. The zero mode integration is supposed to restore the space-time symmetries in a manner similar to the mixed phase at first order phase transition where the position and the orientation of a domain wall are zero modes. We do not follow this rather complicated dynamical process and stop the RG trajectory when such saddle points are encountered.

The bi-local saddle point is generated by an inhomogeneous IR field, $\chi\ne0$, which breaks the space-time symmetries of the functional integration \eq{blocking}. We restrict our attention to such a saddle point without zero mode, $\phi_p$, given by the linearized equation of motion,
\be\label{regsaddl}
\varphi^s=P^{(k-\dk,k)}D[\chi]L,
\ee
where 
\be
P^{(k_1,k_2)}_{x,y}=\int_p\Theta(|p|-k_1)\Theta(k_2-|p|)e^{-i(x-y)p},
\ee
denotes the projector onto the momentum shell of the modes to be eliminated, the source,
\be
L_x=U'(\chi_x)+\int_z[\partial_1V_{x-z}(\chi_x,\chi_z)+\partial_2V_{z-y}(\chi_z,\chi_y)],
\ee
is $\ord{\dk}$ in the $x$-space. The non-linear part of the equation of motion can be neglected in the limit $\dk\to0$. The saddle point \eq{regsaddl} generates the change
\be\label{treetot}
\Delta_kS^{tree}=\hf\int_{xy}\varphi^s_xD^{-1}_{x-y}[\chi]\varphi^s_y,
\ee
in the action which can be simplified by truncating it to our bi-local ansatz,
\be\label{treeev}
\Delta_kS^{tree}=\hf\int_{xy}\bar L_x D^{(k-\dk,k)}_{x-y}\bar L_y,
\ee
where the $\chi$-dependence is ignored in the propagator, $D^{(k-\dk,k)}=P^{(k-\dk,k)}DP^{(k-\dk,k)}$, and the source is simplified to
\be
\bar L_x=U'(\chi_x)+2\int_z\partial_1V_{x-z}(0,\chi_z).
\ee

\begin{figure}
\includegraphics[scale=.5]{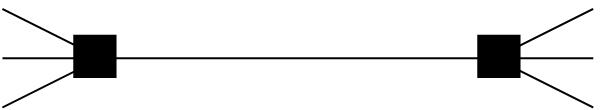}\hskip.5cm\includegraphics[scale=.5]{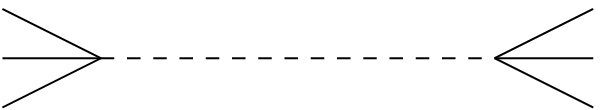}\hskip.5cm\includegraphics[scale=.5]{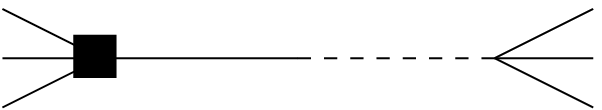}
\caption{The graphical representation of the terms, appearing in the right hand side of the evolution, \eq{bilev}. The solid line denotes the propagator $D^{(k)}$ within the thin shell of modes, being eliminated, the dashed line stands for the bi-local potential, $V_{x-y}(\chi_x,\chi_y)$ and the black square represents $U'(\chi)$.}\label{treef}
\end{figure}

The single cluster part of eq. \eq{treeev} is vanishing hence the local part of the action does not evolve on the tree-level. The evolution of the bilocal part,
\be\label{reeeveqpr}
\Delta_kV^{tree}_q(\chi_1,\chi_2)=2kD^{(k-\dk,k)}_q\left[\partial_2V_q(\chi_1,0)+\hf U'(\chi_1)\right]\left[\partial_1V_q(0,\chi_2)+\hf U'(\chi_2)\right],
\ee
c.f. Fig. \ref{treef}, receives contributions from two different sources: The IR field, $\chi$, generates tree-level renormalization by the help of a vertex $\ord{\phi^n}$ of the local potential $U(\phi)$ if $k\le(n-1)|p|\le\Lambda$ for $n>2$ (the UV modes are not excited by an $\ord{\phi^2}$ vertex owing to the restriction of the $\chi$ within the IR momentum shell) and by the bilocal term $V_q$ with $|q|=k$. The restriction of the function $V_{x-y,m,n}$ to a sixth order polynomial in the second equation of \eq{polynans}, yields the evolution equation,
\be\label{treveqmn}
\Delta_kV^{tree}_{q,m,n}=2\delta_{m,3}\delta_{n,3}D^{(k-\dk,k)}_q\left(V_{q,3,1}+\hf g_4\right)^2.
\ee

\subsection{Fluctuation driven evolution}
The evolution equation \eq{discrev} yields the fluctuation induced contribution,
\be
\Delta_kS^{fl}_k[\Phi+\chi]=-\frac\hbar2\Tr\ln[D^{-1}+\Delta_\Phi\Sigma[\chi]],
\ee
to the evolution equation. In the presence of an arbitrary homogeneous field component, $\Phi$, it is sufficient to retain the $\ord{\chi^2}$ terms on the right hand side,
\bea
\Delta_kS^{fl}_k[\Phi+\chi]&\approx&\frac\hbar2\biggl\{-\Tr\ln D^{-1}+D^{(k-\dk,k)}_{0-0}\int_x\Delta_\Phi U''(\chi_x)\nn
&&+\int_{xy}\bigl(2D^{(k-\dk,k)}_{y-x}\Delta_\Phi\partial_1\partial_2V_{x-y}(\chi_x,\chi_y)\nn
&&+D^{(k-\dk,k)}_{0-0}[\Delta_\Phi\partial^2_1V_{x-y}(\chi_x,\chi_y)+\Delta_\Phi\partial^2_2V_{x-y}(\chi_x,\chi_y)]\bigr)\biggr\}.
\eea
The order $\ord{\chi^0}$ describes the evolution of the local part of the action,
\be\label{wh}
\dot W^{fl}=-\hbar k^d\alpha_d\ln[k^2+m^2+K_k],
\ee
where the dot stands for $k\partial_k$ and $\alpha_d=1/2^d\pi^{d/2}\Gamma(d/2)$, a slight generalization of the Wegner-Houghton equation \cite{wh}. It is easy to check that the $\ord{\chi}$ part of the evolution equation is a consistency condition, the derivative of eq. \eq{wh} with respect to $\Phi$. 

\begin{figure}
\includegraphics[scale=.3]{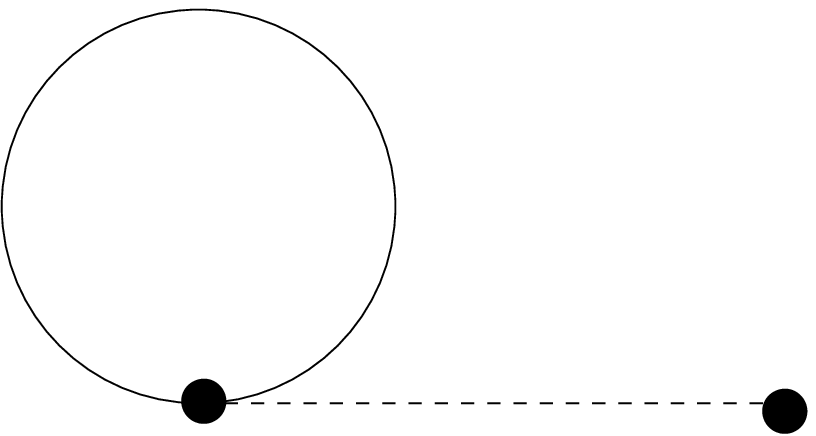}\hskip.5cm\includegraphics[scale=.3]{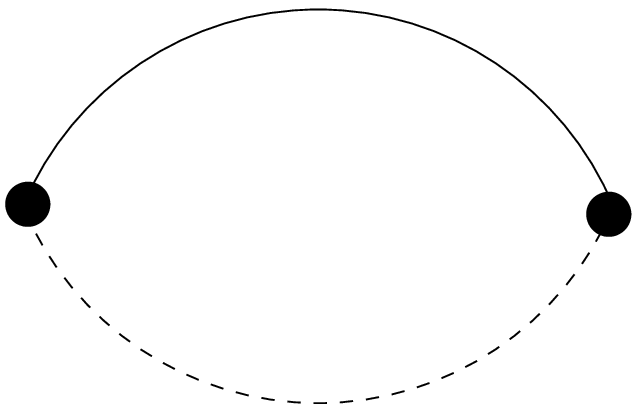}\hskip.5cm\includegraphics[scale=.3]{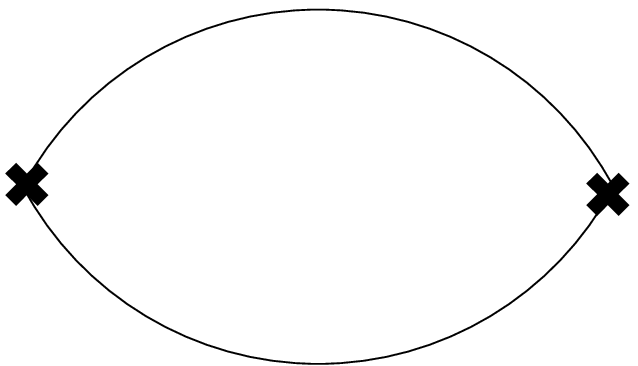}\hskip.5cm\includegraphics[scale=.3]{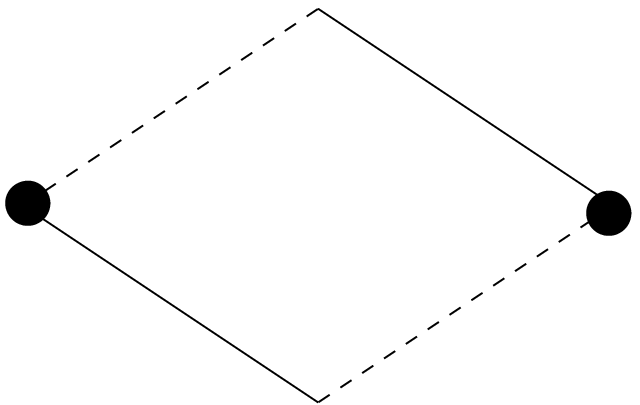}\hskip.5cm\includegraphics[scale=.3]{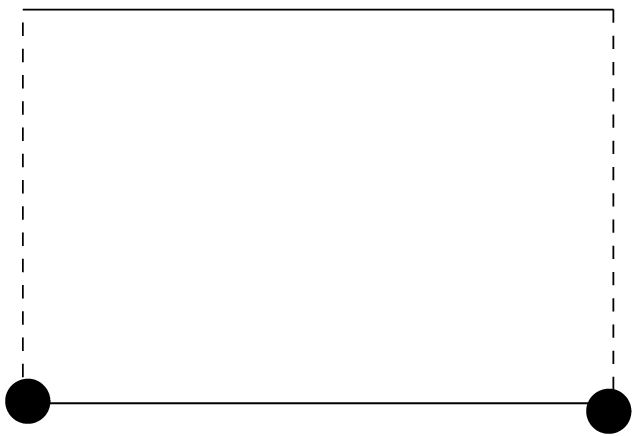}\hskip.5cm\includegraphics[scale=.3]{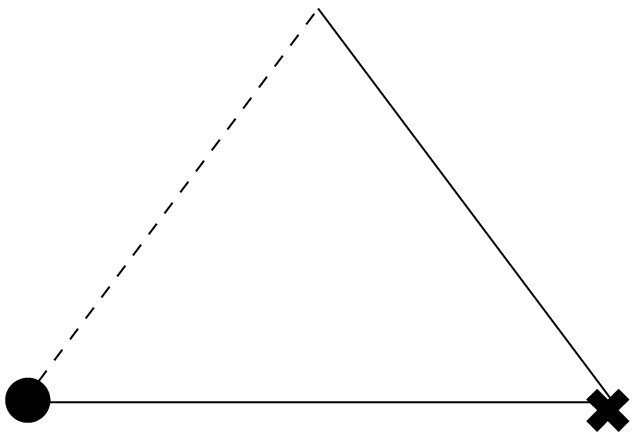}
\caption{The graphical representation of the terms, appearing in the right hand side of the evolution, \eq{bilev}, the external legs suppressed. The notation is the same as in Fig. \ref{treef} except that the cross represents the local vertex, $U''(\chi)$.}\label{loopb}
\end{figure}

The evolution of $V$ is driven by the $\ord{\chi^2}$ bi-local terms,
\bea\label{bilev}
&&\Delta_kV^{fl}_{x-y}(\chi_x,\chi_y)=\nn
&&~~\hbar\biggl\{-\hf D^{(k-\dk,k)}_{0-0}[\Delta_\Phi\partial^2_1V_{x-y}(\chi_x,\chi_y)+\Delta_\Phi\partial^2_2V_{x-y}(\chi_x,\chi_y)]\\
&&~~-D^{(k-\dk,k)}_{y-x}\Delta_\Phi\partial_1\partial_2V_{x-y}(\chi_x,\chi_y)+\frac14D^{(k-\dk,k)}_{x-y}D^{(k-\dk,k)}_{y-x}\Delta_\Phi U''(\chi_x)\Delta_\Phi U''(\chi_y)\nn
&&~~+2\int_{ac}D^{(k-\dk,k)}_{a-x}\Delta\partial_1\partial_2V_{x-c}(\chi_x,0)D^{(k-\dk,k)}_{c-y}\Delta\partial_1\partial_2V_{y-a}(\chi_y,0)\nn
&&~~+2\int_{cd}D^{(k-\dk,k)}_{y-x}\Delta\partial_1\partial_2V_{x-c}(\chi_x,0)D^{(k-\dk,k)}_{c-d}\Delta\partial_1\partial_2V_{y-d}(\chi_y,0)+D^{(k-\dk,k)}_{x-y}\int_a\nn
&&~~\hskip-3pt[D^{(k-\dk,k)}_{a-x}\Delta U''(\chi_x)\Delta\partial_1\partial_2V_{y-a}(\chi_y,0)+\Delta U''(\chi_y)D^{(k-\dk,k)}_{y-a}\Delta\partial_1\partial_2V_{x-a}(\chi_x,0)]\biggr\},\nonumber
\eea
the different terms being represented by the graphs of Fig. \ref{loopb}. The $\ord{D^{(k,\dk)2}}$ contributions are suppressed as $\dk\to0$ and the polynomial ansatz, \eq{polynans}, turns the rest of \eq{bilev} into a set of coupled ordinary differential equations,
\be
\dot V^{fl}_{q,m,n}=-\hbar k\left[\hf(V_{q,m+2,n}+V_{q,m,n+2})\int_pD^{(k)}_p+\int_pD^{(k)}_pV_{p-q,m+1,n+1}\right],
\ee
for $m,n\ge1$ and $D^{(k)}_q=\delta(|q|-k)D_q$. The $O(d)$ invariance allows us to simplify this result to 
\be
\dot V^{fl}_{q,m,n}=-\hbar\frac{k^d}{\omega_k^2}\left[\alpha_d(V_{q,m+2,n}+V_{q,m,n+2})+\frac{\alpha_{d-1}}\pi\int_{-1}^1dc(1-c^2)^\frac{d-3}2V_{\sqrt{q^2-2kqc+k^2},m+1,n+1}\right],
\ee
with $\omega^2_k=m^2+k^2+K_k$. The momentum integral is rather simple for $d=3$,
\be\label{vflqnm}
\dot V^{fl}_{q,m,n}=-\hbar\frac{k^2}{\omega_k^2}\left[\alpha_3k(V_{q,m+2,n}+V_{q,m,n+2})+\frac{\alpha_2}{\pi q}\Theta(\Lambda-k-q)\Theta(k-q)\int_{k-q}^{k+q}dppV_{p,m+1,n+1}\right].
\ee

\subsection{$\ord{\phi^6}$ evolution equations in $d=3$}
We now restrict the potentials \eq{polynans} to $\ord{\phi^6}$ polynomials. The evolution of the off-diagonal elements of $V$ obey the equation,
\be
\dot V_{q,m,n}=-\hbar\frac{k^3}{4\pi^2\omega_k^2}(V_{q,m+2,n}+V_{q,m,n+2}),
\ee
with $\omega_k^2=k^2+g_2+2V_{k,1,1}$ for  $2\le n+m\le6$, $m\ne n$. The evolution of the component $m=n=3$,
\be\label{odvev}
\Delta_kV_{q,3,3}=2D^{(k-\dk,k)}_q\left(V_{q,3,1}+\hf g_4\right)^2,
\ee
contains only the tree-level part and assumes the form of a finite difference equation. The other two diagonal elements, $m=1,2$, satisfy the equation
\be\label{nndvev}
\dot V_{qmn}=\hbar\frac{k^3}{4\pi^2\omega_k^2}\left[\frac1{kq}\int^{k+q}_{|k-q|}dppV_{pm+1n+1}-V_{qm+2n}-V_{qmn+2}\right]
\ee
with $m=n$.

The evolution equation \eq{wh} couples the local potential and $V_{0,m,n}$. Eq. \eq{odvev} together with eq. \eq{nndvev} can be used in the limit $q\to0$ to eliminate $V_{0,m,n}$, yielding
\bea\label{locev}
\dot g_2&=&-\frac{\hbar k^3}{4\pi^2}\frac{g_4}{\omega_k^2},\nn
\dot g_4&=&\frac{\hbar k^3}{4\pi^2}\left(3\frac{g_4^2}{\omega_k^4}-\frac{g_6}{\omega_k^2}\right)-6 \dot{V}_{0,2,2},\nn
\dot g_6&=&\frac{\hbar k^3}{4\pi^2}\left(\frac{15g_4g_6}{\omega_k^4}-\frac{30g_4^3}{\omega_k^6}\right),
\eea
with $\omega_q^2=q^2+g_2+2V_{q,1,1}$.

\section{Phase diagram and renormalized trajectories}\label{sec:dis}
We turn to the phase diagram and the renormalized trajectories of the model. First we discuss the possible relevant bi-local coupling constants and clarify the order parameter, used to map the phase diagram. Next we describe the tree-level solution, followed by the presentation of the full, fluctuations driven flow diagram. 

The peculiarity of the tree-level evolution equation \eq{reeeveqpr} is that $V_q$ is touched only at $k=|q|$, there is no accumulation of contributions of the blocking at given $q$. Nevertheless the $q$-dependence can be used to define tree-level scaling operators, $V_{q,m,n}=V_{\ell mn}|q|^{-\ell}$, with mass dimension $[V_{\ell,m,n}]=d-(m+n)(\frac{d}2-1)+\ell$. The clusterization of the action for a homogeneous field configuration imposes the limit $\ell<d$. Let us consider a pair of field powers, $m$ and $n$, such that the maximum of $[V_{\ell,m,n}]$ is positive, $m+n\le4d/(d-2)$, then the coupling constant $V_{\ell,m,n}$ is marginal for $\ell=\ell^*=(m+n)(d-2)/2-d$ and relevant for $\ell^*<\ell<d$. The negative integer values of $\ell$ correspond to the usual local vertices and the remaining spectrum yields the truly non-local relevant and marginal bi-local operators. In the present case, $d=3$, the vertices $2\le n+m\le12$ are marginal for $\ell^*=(m+n-6)/2$ and relevant for $(m+n-6)/2<\ell<3$. The loop-corrections generate and $\ord\hbar$ modification of the critical exponents but this is important in the marginal case only, the tree-level relevant or irrelevant operators do not change their class. The lesson of this simple argument is that we may encounter new relevant operator compared to the local model.

The scale-dependence is better described by the bi-local potential than the local one hence it is better be cautious and look for possible order parameters, $\la\phi_x\ra=\phi\cos(px^1)$, at an arbitrary momentum scale. The parameters $\phi$ and $p$ are determined by a simple, tree-level minimization of the action density,
\bea\label{actdens}
s(\phi,p)&=&\frac1V\left[\hf\int_x\phi_xD_0^{-1}\phi_x+\int_xU(\phi_x)+\int_{xy}V_{x-y}(\phi_x,\phi_y)\right]\nn
&=&\frac{p^2\phi^2}4+s_1+s_2,
\eea
where $V=\int_x1$ and
\bea
s_1&=&\sum_n\frac{g_{2n}\phi^{2n}}{2^{2n}(n!)^2}\nn
s_2&=&\sum_{mn}\frac{\phi^{m+n}}{m!n!2^{m+n}}\sum_{j=\max(0,\frac{m-n}2)}^{\min(m,\frac{m+n}2)}\left(^m_j\right)\left(^n_{\frac{m+n}2-j}\right)V_{(2j-m)p,m,n}.
\eea
The action density, truncated at $\ord{\phi^6}$, can be written in the form
\be\label{semien}
s(\phi,p)=a(p)\phi^2+\frac{b(p)}2\phi^4+\frac{c(p)}{3!}\phi^6,
\ee
with
\bea
a(p)&=&\frac{p^2+m^2+K_p}4,\nn
b(p)&=&\frac18\left(\frac{g_4}4+\frac{V_{2p,2,2}}2+2V_{p,3,1}+V_{0,2,2}\right),\nn
c(p)&=&\frac3{32}\left(\frac{g_6}{36}+\frac{V_{3p,3,3}}{18}+\frac{V_{p,3,3}}2\right).
\eea
The stability requires the condition $c>0$, the action displays non-trivial minima with $\phi\ne0$ if $d=b^2-8ac/3$ is positive and the minimum of the action is reached at the maximum of $d$.

\subsection{Tree-level phase diagram}
The tree-level evolution, discussed in section \eq{treeevs}, keeps the local potential unchanged, $U_k(\phi)=U_\Lambda(\phi)$, and the bi-local potential, the solution of eq. \eq{reeeveqpr},
\be\label{treertrbil}
V^{tree}_{kx-y}(\chi_1,\chi_2)=-\hf U'(\chi_1)\hD^{(k,\Lambda)}_{x-y}U'(\chi_2),
\ee
preserves the analytic structure as long as the action functional is convex. We have $K_q=0$ for $|p|<k$ hence that the dependence of the action on the homogeneous IR field component, $\Phi$, is concave for $k^2<-g_2$. Hence the symmetry breaking saddle point with zero modes appears when $g_2<0$ and we stop the evolution at $k^2=-g_2$. Another reason to stop the evolution is the loss of stability, the boundedness of the action from below, indicated by the inequality $c(p)<0$ for some $|p|<k$. 

\begin{figure}
\includegraphics[width=8cm,angle=-90]{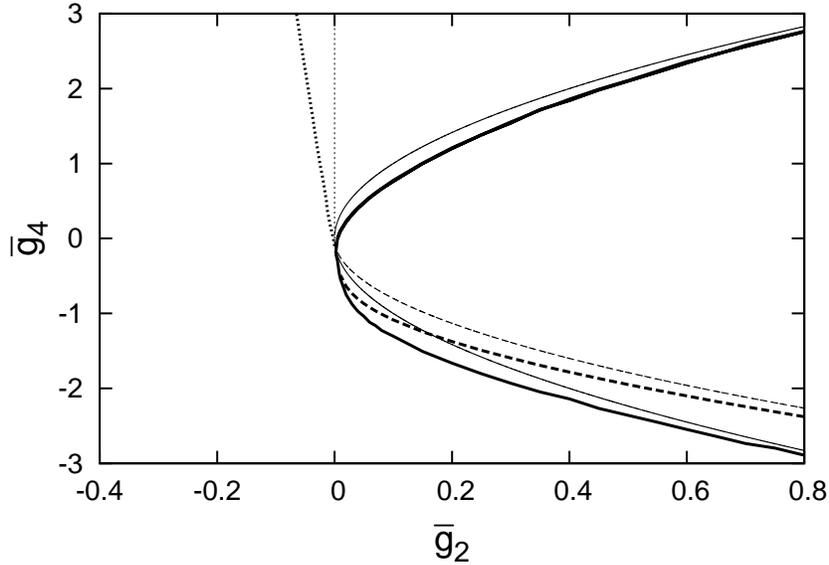}
\caption{\label{fig:phsstr} The phase structure on the $(\bar g_2,\bar g_4)$ plane at $\bar g_6(\Lambda)=10$. The tree- and the loop-level phase boundaries are depicted by thin and thick lines, respectively. The vertical dotted lines indicate a second order phase transition for $g_4>0$. The bi-local theory is stable, $c>0$, within the horizontal parabola, $|h|=1$, shown by a solid line. The dashed lines represent the tree-level first order transition. The fluctuations restore the symmetry for weakly negative $g_2$ and induce a small shift by decreasing $g_4$ for positive $g_2$.}
\end{figure}

It is well known that the interplay of $\phi^4$ and $\phi^6$ terms may induce a first order phase transition. However the tree-level $\ord{\phi^6}$ bi-local term, generated by the $\phi^4$ term of the local potential, is attractive hence we assume a sufficiently large $g_6>0$ in the initial action to explore such a phase transition.

The phase structure is depicted in Fig. \ref{fig:phsstr} on the plane $(\bar g_2,\bar g_4)$ of the dimensionless coupling constants, $\bar g_n=g_n\Lambda^{(n-6)/2}$ at $\bar g_6=10$. It is instructive to consider first the phase structure of the local model, $V=0$, when the stability, $c>0$, requires $g_6>0$. The vacuum is trivial and the order parameter is vanishing, $\phi_x=0$, in the symmetric phase with $g_2,g_4>0$. The order parameter becomes non-vanishing, $\phi_x=\Phi\ne0$, in a continuous manner as $g_2$ changes sign, signaling a second order phase transition at the quadrant plane $g_4,g_6>0$ at $g_2=0$, shown by the vertical dotted line. The trivial vacuum extends slightly over negative values of $g_4$ if $g_2>0$ and ends at a first order phase transition line, $d=0$, denoted by thick continuous line in the figure. A homogeneous condensate is formed for $d>0$, within the region $g_2>0$ and $h=g_4/\sqrt{g_2g_6}=\bar g_4/\sqrt{\bar g_2\bar g_6}<-4/3$. Note that the condensate, indicating the spontaneously broken $\phi\to-\phi$ symmetry, is homogeneous since the kinetic energy is positive definite. This is the mean-field phase diagram which is modified by the tree-level contributions of the evolution equation, the symmetry breaking saddle point with zero modes, \eq{singlsp}, realizing the generalized Maxwell-construction and rendering the action convex or degenerate \cite{tree}. 

The inclusion of the bi-local part of the action modifies the term $\ord{\phi^6}$ of the action density \eq{semien} which influences neither the emergence of the saddle points with zero modes nor the second order phase boundary however makes the theory unstable for $g_2>0$ and $|h|>1$, both in the symmetric and the first-order transition realized symmetry broken phase. The border of the stable region for $g_2>0$ is indicated by the thick solid line in the figure. The discriminant, $d(p^2)$, is a non-increasing function hence the oscillatory vacuum, if exists, is reached at $|p|=k/3$. Hence the oscillatory vacuum is preferred below the dashed line, for $d(k^2/9)>0$ or $h<-4/5$. The homogeneous vacuum configuration has higher action density than the oscillatory vacuum in the stable theory.

\subsection{Fluctuations induced phase diagram}
The local coupling constants change according to eq. \eq{locev} and the bi-local coupling functions, $V_{q,m,n}$ follow eqs. \eq{odvev}-\eq{nndvev}. The tree-level evolution of $V_{q,3,3}$ is generated by a graph where the UV mode propagator connects two fourth order local vertices. The three outer legs, joining at the same vertex, produce the momentum $|p|=k$, needed for the UV excitation, and ignites the evolution of the other bi-local coupling functions via eq. \eq{nndvev}. The evolution of the bi-local coupling functions is influenced by the local coupling constants and the last term in the beta function of $g_4$ in \eq{locev} represents the reaction of the bi-local part of the action to the local one in the evolution. The numerical results below are given in units $\hbar=1$.

We mention four modifications of the phase diagram in the coupling constant space $(\bar g_2,\bar g_4,\bar g_6)$ due to the fluctuations: 

(i) The fluctuations deform the phase boundaries. This is a simple quantitative shift for medium $g_6$, shown in Fig. \ref{fig:phsstr}, the second order phase transition surface is slightly bended towards more negative values of $\bar g_2$ and $\bar g_4$ and the first order transition and the instability surfaces are also shifted downwards along the $\bar g_4$ axes. The change is more pronounced for weak or strong $g_6$. The second order phase transition surface, found for $g_2<0$ in the tree-level solution, is deformed in a substantial manner and the bi-local fluctuations driven instability seems to eliminate the symmetry breaking at weak $g_6$, c.f. Fig. \ref{fig:lpgbl} (b) below. The first order transition driven ordered phase structure changes in a qualitative manner at $g_2>0$ for strong $g_6$. A new phase boundary can be seen, c.f. point (iv) below and our exploratory numerical work can not exclude further phase boundaries.

(ii) The fluctuation induced evolution of the local coupling constants display a non-Gaussian fixed point, reminiscent of the Wilson-Fisher fixed point of local theories. There is though an important difference with respect to the local case, namely only the local coupling constants become scale invariant at that point, separating phases with and without a condensate, and the bilocal parameters of the action continue their fluctuations induced evolution. The non-Gaussian partial fixed point lies very close to the instability surface, within the stable domain. 

(iii) The semi-infinite plane, $\bar g_2=0$, $g_4<0$ cuts the broken symmetric phase of the tree-level theory into two parts, the condensate being generated by first or second order phase transition for $\bar g_2>0$ and $\bar g_2<0$, respectively. One expects that a deformed surface remains present in the phase diagram of the full solution, including the fluctuations, and represents a first order phase ordered-ordered transition, defined by a discontinuity of the order parameter, assuming non-vanishing values on both side. 

(iv) The fluctuations induce another surface, separating two qualitatively different regions within the first order phase transition induced broken symmetric phase with smaller or larger values of $\bar g_2$. The semiclassical vacuum is oscillatory with $|p|=k/3$ for $k>k_{min}>0$ for the former and down to the IR end point, $k_{min}=0$, in the latter case. 

The modulated vacuum has qualitatively different physics than the homogeneous one. The translation invariance, broken in an infinite volume, is not recovered and the excitation spectrum splits into different dispersion relations, in a manner similar to solids. The wavelength $\lambda$ of the modulation of the vacuum runs with the cutoff at large $g_2$, $\lambda\sim1/k$, i.e. the vacuum seems to be homogeneous in any finite space region. However there is no gap in the excitation spectrum owing to the acoustic phonons, the Goldstone modes of the broken external symmetry. The homogeneity of the vacuum, suggested by the minimization of our simple ansatz, \eq{actdens}, for weak $g_2$ at low $k$, has different interpretations. One the one hand, it may indicate that the translation invariance of the vacuum is truly recovered at sufficiently long distances. On the other hand, it might be the result of the oversimplified structure of the ansatz \eq{actdens}. Consider for instance a simple construction of the block spin, using the majority rule for neighboring spins in a spin model with in an anti-ferromagnetic Neel ground state where the anti-ferromagnetic order is  destroyed by a single inappropriate blocking. In a similar manner, a theory in a vacuum, modulated with some wave vector, $\v{p}$, may represent an IR Landau pole of the RG trajectories at $k=|\v{p}|$ since a simple ansatz may not reproduce the impact of the modulation of the vacuum for $k<|\v{p}|$. A further level of complexities, in interpreting the modulated phase, arises from the simplicity of our plane wave ansatz of the semiclassical vacuum, detecting the breakdown $O(d)\to O(d')$, $d'=d-1$, of the external symmetry. One expects more involved symmetry breaking patterns, $d'<d-1$, as well \cite{afvacuum} and the inhomogeneous phase may include further phase boundaries.

\subsection{Renormalized trajectories}
\begin{figure}
\includegraphics[width=8cm,angle=-90]{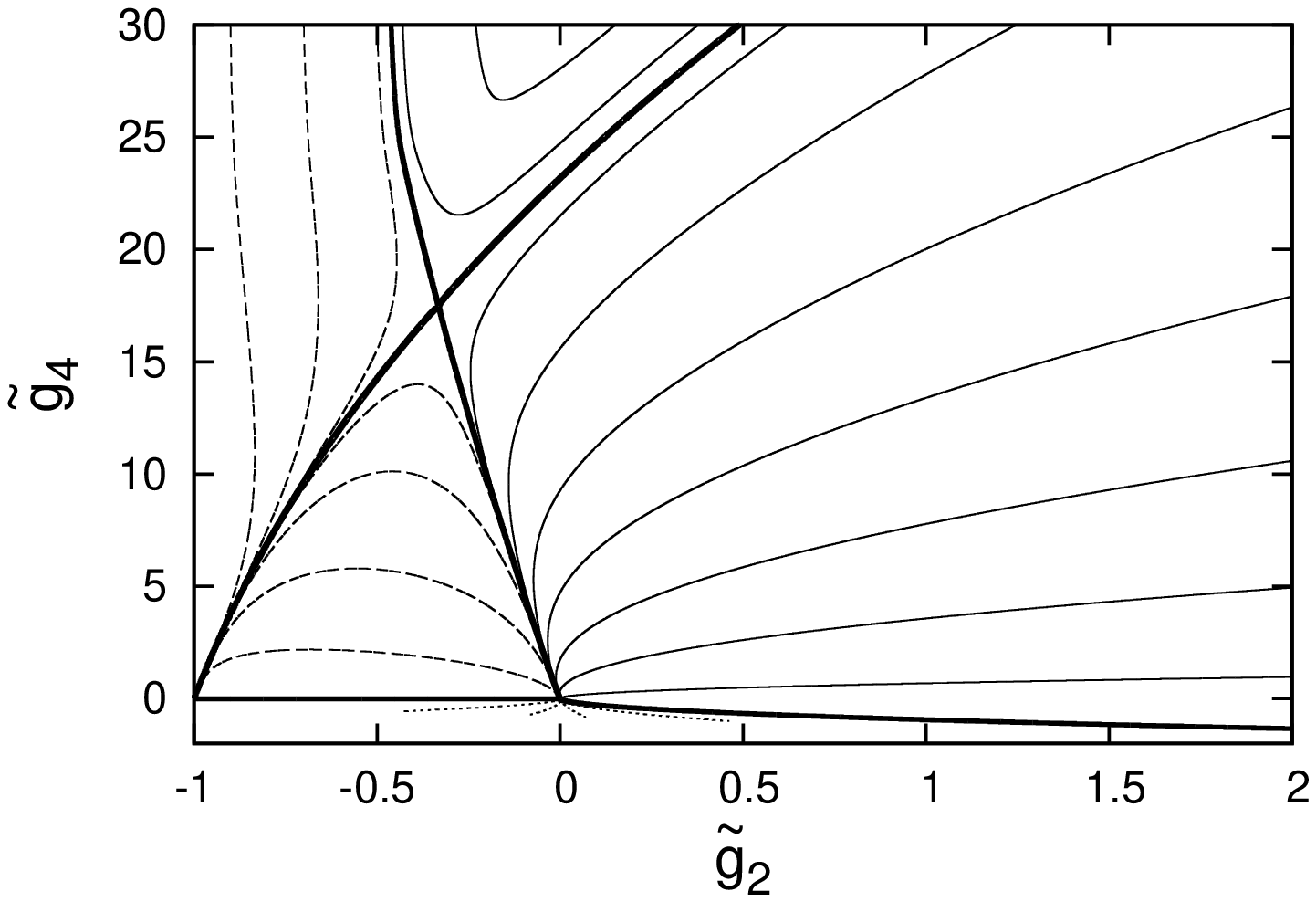}

\hskip.5cm(a)

\includegraphics[width=8cm,angle=-90]{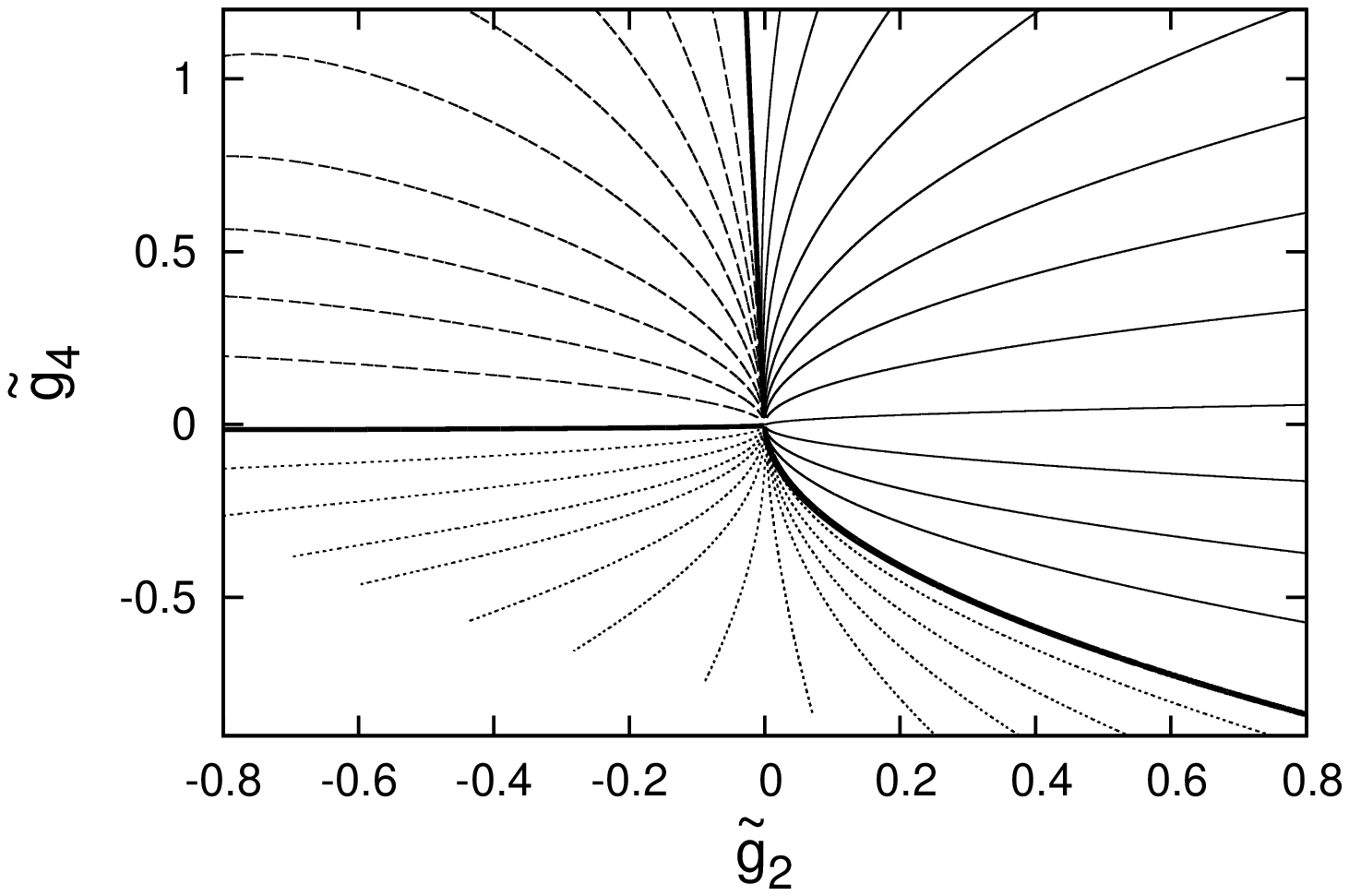}

\hskip.5cm(b)
\caption{\label{fig:ltr} (a): The projection of few renormalized trajectories with $g_6(\Lambda)=0.1$ on the plane $(\tilde g_2,\tilde g_4)$ of the local model, (b): zoom around the Gaussian fixed point. The solid curves run within the  symmetric phase and the dashed curves belong to the symmetry broken phase, generated by a second order phase transition. The evolution is stopped at a finite value of the cutoff where a symmetry breaking saddle point with zero mode is generated. The trajectories, shown by the dotted lines, stop at the instability, $c=0$.}
\end{figure}

The flow diagram of the renormalized trajectories offers another view of the phase diagram, organized in terms of the running coupling constants, $\tilde g_n=g_nk^{(n-6)/2}$, whose dimension is removed by the gliding cutoff. 

Let us start with the local model which supports two fixed points, the Gaussian, $\t g^*_{G2}=\t g^*_{G4}=\t g^*_{G6}=0$ and the Wilson Fisher (WF),
\be
\t g_{WF2}^*=-\frac13,~~~
\t g_{WF4}^*=\frac{16}9\frac{\pi^2}\hbar,~~~
\t g_{WF6}^* = \frac{256}{27}\left(\frac{\pi^2}\hbar\right)^2.
\ee
The critical exponents of the linearized blocking at the Gaussian fixed points, $\tilde g_{sc}\approx k^\nu$, with $\nu_G=\{-2,-1,0\}$, correspond to the scaling combinations of the coupling constants, $\tilde g_{sc}$, are dominantly $g_2$, $g_4$ and $g_6$, respectively. The critical exponents, $\nu_{WF}=\{19.34,-1.71,1.36\}$, of the WF fixed point belong to scaling combinations which are dominated by $g_6$. The vanishing of the relevant scaling combinations at a fixed point defines separatrix surfaces which extend into the non-asymptotic scaling regions, as well, representing phase boundaries. 

The projection of the trajectories is shown on the $(\tilde g_2,\tilde g_4)$ plane in Fig. \ref{fig:ltr} (a). The almost horizontal thick line from the Gaussian fixed point with $\tilde g_2>0$ runs along the first order phase transition where the trajectories within the symmetry broken phase run into instability, $\t g_6<0$. This phase may need higher order terms in $\phi$ in the bare potential to stabilize. The thick separatrix, going upward from the Gaussian fixed point represents a second order phase transitions. The two irrelevant coupling constants of the Wilson-Fisher fixed point make the trajectories converge to the singular point, $(\tilde g_2,\tilde g_4)=(-1,0)$, or to $(\tilde g_2,\tilde g_4)=(\infty,\infty)$, in the broken symmetric and the symmetric phase, respectively, on this projected plane. The thick solid line, connecting the singular and the Gaussian fixed points separates regions of the symmetry broken phase where the order parameter is generated by first and second order transitions. The phase transition surfaces are approximately orthogonal to the coupling constant axes around the Gaussian fixed point, simplifying the RG flow, shown in Fig. \ref{fig:ltr} (b). 

\begin{figure}
\includegraphics[width=8cm,angle=-90]{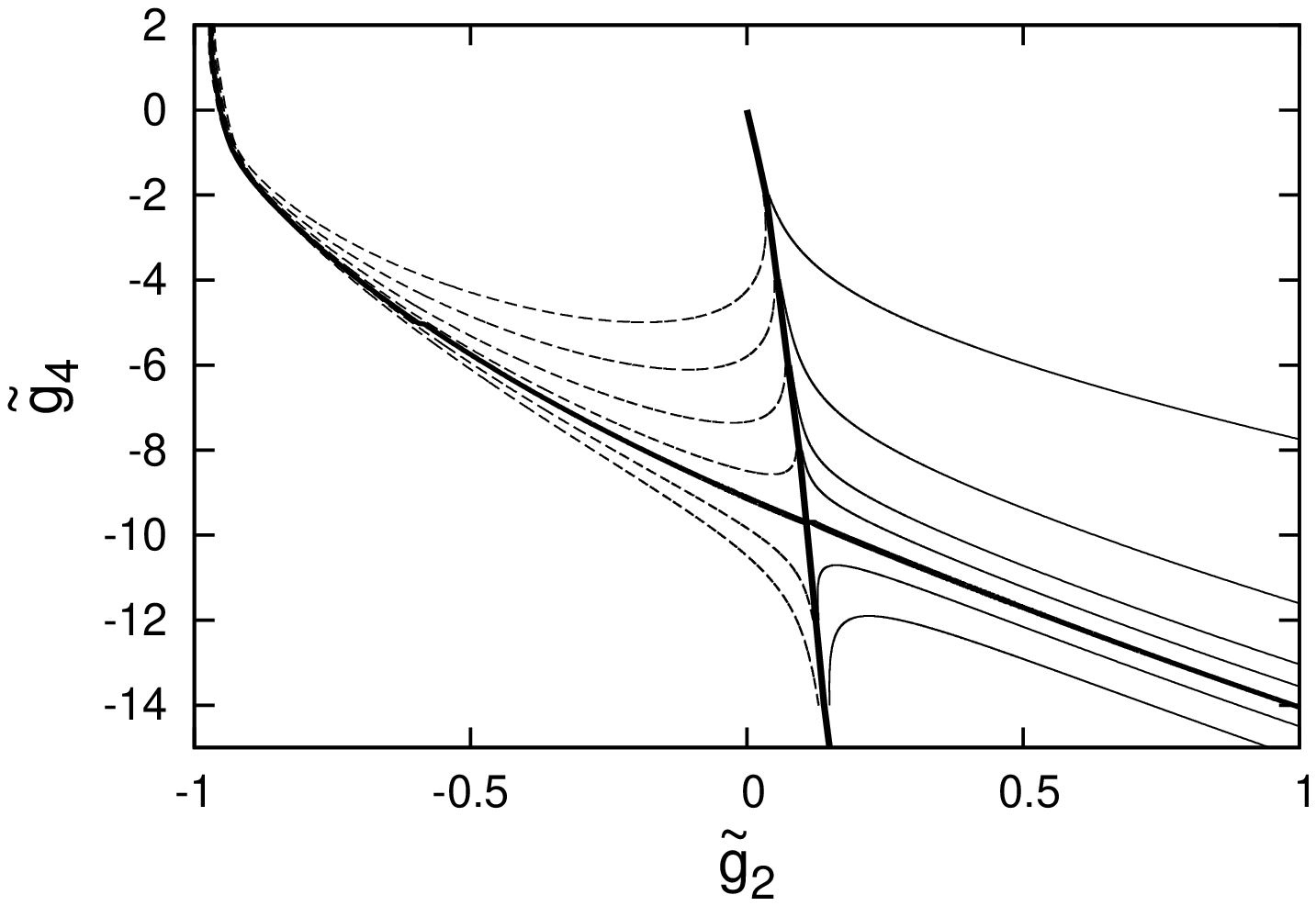}

\hskip.5cm(a)

\includegraphics[width=8cm,angle=-90]{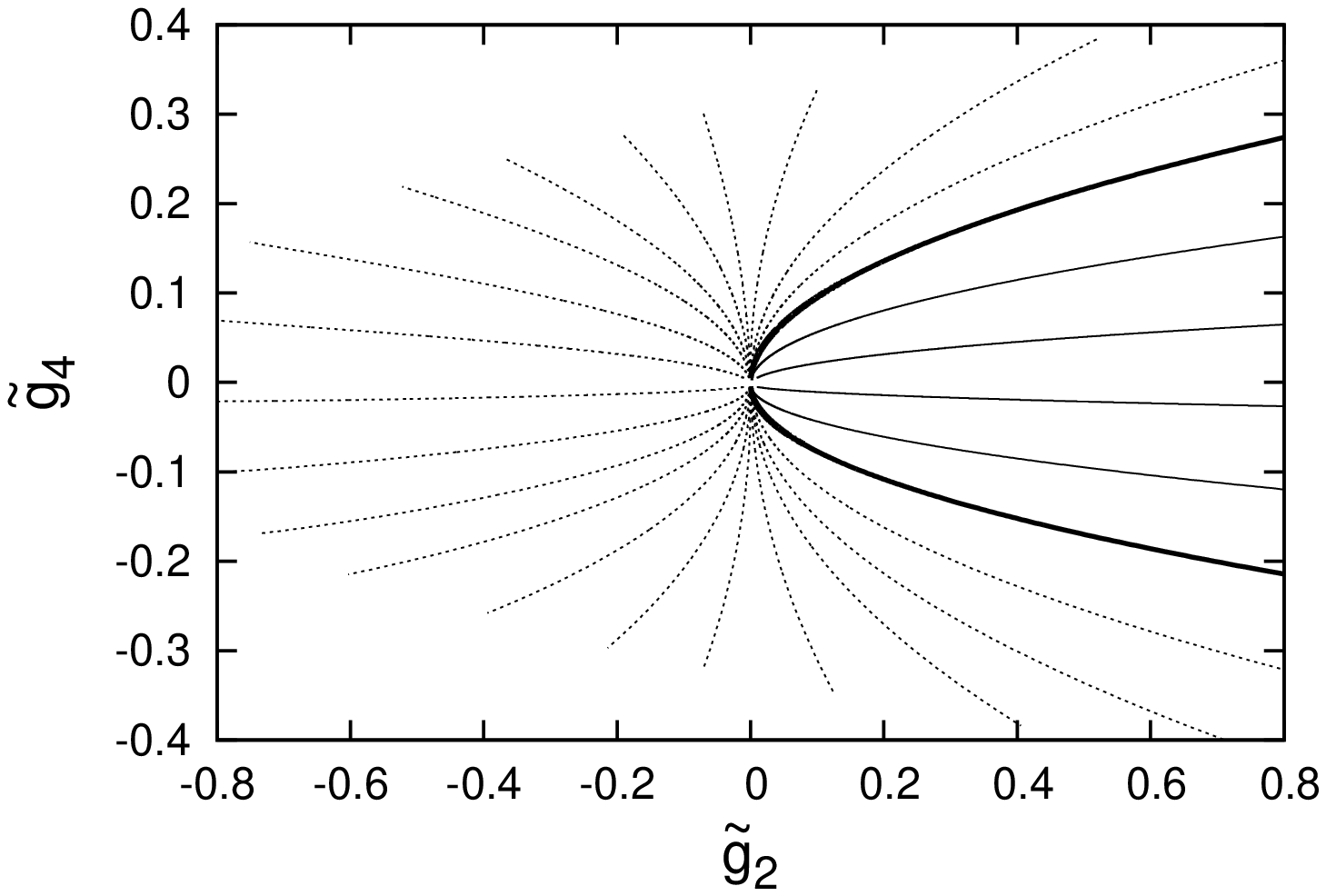}

\hskip.5cm(b)
\caption{\label{fig:lpgbl} The same as Fig. \ref{fig:ltr} except for the bi-local theory.}
\end{figure}

While the bi-local terms do not change the linearized scaling laws of the local coupling constants around the Gaussian fixed point they do change the RG flow at stronger coupling strengths in a considerable manner, namely both the non-Gaussian partial fixed point and the singular point are displaced on Fig. \ref{fig:lpgbl} (a). The partial fixed point is now at
\be
\t g_2^*=\frac19,~~~
\t g_4^*=-\frac{80}{81}\frac{\pi^2}\hbar,~~~
\t g_6^*=\frac{1280}{729}\left(\frac{\pi^2}\hbar\right)^2,~~~
\t v_{022}=\frac{400}{41^2}\frac{\pi^2}{\hbar^2}
\ee
where the scaling combinations of the local coupling constants remain dominated by $\tilde g_6$ and acquire the exponents $\nu=\{-3.63,-2.08,0.71\}$. Both relevant scaling operators are dominated by $\tilde g_6$, bending strongly the phase transition surfaces and reducing the second order phase transition generated symmetry broken phase considerably. The shaded lines, the projection of the RG trajectories of this phase on the $(\t g_2,\t g_4)$ plane in Fig. \ref{fig:lpgbl} (a) were found by careful fine tuning the initial conditions as opposed to the solid lines, the trajectories of the symmetric phase.  The dashed and the solid line trajectories are separated by the new relevant operator and the upper (lower) region, corresponding to larger (smaller) $\t g_6$, belongs to the symmetric (broken symmetric) phase. The instability, induced by the bi-local terms, wipes out the broken symmetric phase, generated by the second order phase transition around the Gaussian fixed point, according to Fig. \ref{fig:lpgbl} (b).

\section{Stability}\label{stabs}
We address finally an unexpected change of the dynamics, namely the bi-local saddle point may destabilize the theory. The point is that the $\ord{\dk^0}$ contribution, retained in section \ref{treeevs}, represents an attractive interaction owing to the inequality,
\be
\frac{\int d\varphi e^{-\frac{a}2\varphi^2+b\varphi}}
{\int d\varphi e^{-\frac{a}2\varphi^2}}=e^{\frac{b^2}{2a}}>1,
\ee
holding if $a>0$, $b\ne0$ because the maximum of the exponent in the numerator is positive. In a more physical terms, the taking into account of correlations, generated by the eliminated modes, can only lower the free energy. For instance a local $\phi^4$ theory looses stability for large field, $\phi^2>\phi^2_{inst}\sim1/g_4$, when the attractive forces of the  bi-local $g_4^2\phi_x^3\phi_y^3$ vertices are stronger than the repulsion, exerted by the local $g_4\phi^4_x$ interactions. The $\ord{\dk}$ tree-level graph contains repulsive contributions, for instance in the leading order $\chi$-dependence in \eq{treetot}
\be
-\hf\int_{xyz}\varphi^s_xD_{x-y}\Delta_\Phi U''(\chi_y)D_{y-z}\varphi^s_z,
\ee
for a convex local potential may stabilize the vacuum for $\phi^2>\phi^2_{stab}\sim\phi^2_{inst}/\dk$.

It is instructive to confront this state of affairs with the lattice regularized $\phi^4$ theory whose stability follows from the boundedness of the potential energy from below as the function of the lattice field variable, $\phi_x$. The instability, encountered in this work, arises by the truncation of the blocked action at $\ord{\dk^0}$. If the theory is defined within a quantization box of size $L$ and is regulated by a momentum-space cutoff then the truncation of the blocked action can be justified only in the thermodynamic limit. For a large but finite size system $\dk=\ord{1/L}$ and the vacuum of the momentum space regulated theory is stable and contains field fluctuations $\phi^2=\ord{L}$, generated by the bi-local terms. In other words, the saddle point changes the usual, $\phi=\ord{L^0}$ vacuum to another one with strong, $\phi=\ord{L^{1/2}}$ field and makes the thermodynamic limit of the sharp momentum cutoff scheme non-continuous. 

The instability of the $\phi^4$ theory appears as a violation of the universality in four dimensions. The bi-local structure of the saddle-point, appearing in infinite systems, introduces new relevant operators, not considered if the theory, defined at the initial cutoff $k=\Lambda$, is local. The non-local interaction vertices are simply ignored in the RG literature altogether. The saddle-point generated instability, an amplification of the role these terms of the blocked action play, can be eliminated by stabilizing the weak field vacuum by a sufficiently strong, local $\ord{\phi^6}$ vertex in the bare theory. In fact, a large enough $g_6$ saves the stability and suppresses the non-commutativity of the thermodynamic limit and the lowering of the cutoff. 

The scalar model, defined by a bare, local action containing vertices up to the marginal order and regulated by the momentum cutoff in infinitely large space is unstable. The perturbative universality can be recovered in dimensions 4 or 5 by introducing a sixth order irrelevant vertex with cutoff-independent dimensionless coupling, $\t g$. In fact, the power counting detects no new divergences in such a theory and the perturbative removal of the cutoff can be carried out as in the case of other regulator. Such a rescue operation of the universality can be justified by requiring infrared stability, a continuous thermodynamic limit.

\section{Summary}\label{sums}
The phase structure of the $\ord{\phi^6}$ scalar model is investigated in three Euclidean dimensions by retaining the bi-local saddle point contribution to the RG flow. The non-local extension of the action shifts the phase boundaries, produces new phases, generates new relevant operators and restricts the universality classes. The new phase appears when the blocked theory develops a modulated vacuum, the relevant operators are bi-local and the restriction of the universality class applies to four and five dimensional $\phi^4$ and $\phi^6$ models, respectively.

These results open several interesting questions, to mention but few. The instability, driven by the bi-local saddle point, restricts drastically the available parameter space of the model. It is a question of central importance to find out the possible increases of the stable region by allowing higher powers of the field in the ansatz. Furthermore, the phase structure is found to be rather involved and more work is needed to determine it in a more detailed manner. In particular, a more reliable determination of the breaking patters of the external symmetry and the distinction between the scenarios of an IR Landau pole or a homogeneous vacuum are needed. The inclusion of bi-local terms in the initial action enlarges the phase space considerably. Another important extension of the present work is the identification of the relevant bi-local operators to map out new universality classes. The comparison of different regulators is specially useful in this respect. Yet another issue, the inhomogeneous saddle point brings up, is that the blocked theory exists in integer dimensions only, casting doubt on the consistency of the $\epsilon$-expansion, defined around the upper critical dimension.

Finally we mention that the bi-local saddle point is supposed to play a more important role in real than in imaginary time where the mass-shell modes, a fundamental element of any realistic model, exist. The outgoing radiation is an essential element of any effective theory and makes the generalization of the present work necessary for real time field theories.

\section*{ACKNOWLEDGMENTS}
S. Nagy acknowledges financial support from a J\'anos Bolyai Grant of the Hungarian Academy of Sciences, and the Hungarian National Research, Development and Innovation Office NKFIH (Grant Nos. K112233, KH126497).

\end{document}